\documentclass[reprint,secnumarabic,amssymb, nobibnotes, aps, prl]{revtex4-1}

\usepackage[utf8]{inputenc}
\usepackage[T1]{fontenc}
\usepackage{graphicx}

\usepackage{siunitx}
\usepackage{hyperref}

\begin{document}

\title{Planar GHz Resonators on SrTiO$_3$: Suppressed Losses at Temperatures below 1~K}

\author{Vincent T. Engl, Nikolaj G. Ebensperger, Lars Wendel, and Marc Scheffler}
\affiliation{1. Physikalisches Institut, Universit\"at Stuttgart, 70569 Stuttgart, Germany}
\date{\today}

\begin{abstract}

The complex dielectric constant $\hat{\epsilon} = \epsilon_1 + i \epsilon_2$ of SrTiO$_3$ reaches high values $\epsilon_1 \approx 2*10^{4}$ at cryogenic temperatures, while the dielectric losses ($\epsilon_2$) are much stronger than for other crystalline dielectrics. SrTiO$_3$ is a common substrate for oxide thin films, like the superconducting LaAlO$_3$/SrTiO$_3$ system, but the large $\epsilon_1$ and $\epsilon_2$ restrict high-frequency quantum devices on SrTiO$_3$.
Here we present superconducting coplanar Nb resonators on SrTiO$_3$, which we successfully operate in a distant-flip-chip geometry at frequencies that exceed 1~GHz. We find a pronounced and unexpected increase in resonator quality factor $Q$ at temperatures below 1~K, reaching up to $Q \approx 800$. We attribute this to substantial changes of the dielectric losses in SrTiO$_3$ at mK temperatures, and we also detect non-monotonous changes in the temperature-dependent $\epsilon_1$.
These findings challenge our present understanding of the dielectric properties of SrTiO$_3$ and at the same time demonstrate that cryogenic high-frequency devices on SrTiO$_3$ are more feasible than previously assumed.
\end{abstract}

\maketitle

When SrTiO$_3$ (STO) is cooled down, its dielectric constant $\hat{\epsilon} = \epsilon_1  + i \epsilon_2$ increases drastically, from already substantial room-temperature $\epsilon_1 \approx 300$ to values beyond $2 \times 10^{4}$. 
Eventually $\epsilon_1$ levels off below 3~K due to quantum fluctuations that suppress an incipient ferroelectric transition \cite{Weaver1959,Mitsui1961,Sawaguchi1962,Sakudo1971,Neville1972,Mueller1979,Viana1994}, and exhibits a weak maximum near 2~K \cite{Fischer1985,Mueller1979,Viana1994,Rowley2014}. These dielectric properties also depend on sample strain and isotope composition as well as on probing frequency and strength of the applied electric field \cite{Mueller1979,Fischer1985,Viana1994,Rowley2014,Davidovikj2017}.
Large values and tunability of $\epsilon_1$ are of interest for electronic circuitry, but there the substantial dielectric losses of STO, quantified by $\epsilon_2$ and loss tangent $\tan \delta = \epsilon_2/\epsilon_1$, are detrimental \cite{Tagantsev2003,Kong2010}, and the large $\epsilon_2$ of STO persists to cryogenic temperatures where other crystalline dielectrics feature very low losses \cite{Krupka1994}.

STO is a common substrate for oxide thin-film growth and prerequisite for the LaAlO$_3$/SrTiO$_3$ (LAO/STO) interface that hosts a highly mobile two-dimensional electron system \cite{Ohtomo2004}. Below 300~mK, LAO/STO even features a gate-tunable superconducting dome \cite{Caviglia2008}, which relates to the superconducting state of doped bulk STO \cite{Lin2013,Stucky2016,Swartz2018,Thiemann2018a,Collignon2019}. 
Superconducting LAO/STO is key ingredient for oxide quantum circuits \cite{Coll2019}, which seek to combine the success of mesoscopic quantum devices fabricated from conventional materials \cite{Eckstein2013} with the rich material properties found in oxides \cite{Coll2019,Lorenz2016,Boschker2017,Dey2018}, and several fundamental elements of quantum electronics have already been demonstrated in LAO/STO \cite{Cheng2015,Goswami2016,Tomczyk2016,Monteiro2017,Thierschmann2018,Pai2018a,Pai2018b}. While these were accessed with dc signals, more advanced solid state quantum circuitry is usually operated at GHz frequencies \cite{Frunzio2005,Gu2017}, which is also the appropriate electrodynamic regime for low-temperature spectroscopy \cite{Driessen2012,Scheffler2013,Thiemann2018b}. 
Microwave devices on STO are difficult to implement, and only very recently high-frequency experiments on superconducting LAO/STO interfaces succeeded: Singh \textit{et al.}\ used mm-sized LAO/STO samples as discrete elements in resonant circuits which operate at a few hundred MHz \cite{Singh2018,Singh2019}. 
Manca \textit{et al.}\ structured a coplanar waveguide resonator into the LAO/STO interface, and they characterized the superconducting state of this one-dimensional resonator at up to the tenth harmonic, at frequencies up to 340~MHz \cite{Manca2019}. 
These studies demonstrate that superconducting resonators based on LAO/STO can be successfully operated, but at the same time they illustrate the fundamental challenges: firstly, the high $\epsilon_1$ of STO leads to rather low fundamental frequencies of resonators with conventional design and makes it difficult to interface them with standard microwave circuitry with 50~$\Omega$ characteristic impedance. Secondly, the quality factors $Q$ of such transmission line resonators so far are orders of magnitude smaller than what has been achieved in planar superconducting resonators on other substrates \cite{Megrant2012}. Resonator $Q$ can be limited by Ohmic losses in the conductor as well as dielectric losses in the substrate, and therefore we now explicitly focus on losses due to the STO. Davidovikj \textit{et al.}\ previously explored this issue \cite{Davidovikj2017}, using superconducting coplanar MoRe resonators on STO. At 3.5~K they observed $\tan \delta \approx 1/Q$ as low as $10^{-3}$, but only for frequencies below 100~MHz. Considering that $\tan \delta$ of STO increases strongly with frequency \cite{Davidovikj2017,Rupprecht1962}, one expects that $Q$ for coplanar resonators on STO operating at 1~GHz or higher will hardly exceed 100. Therefore, our results come as a surprise: for a superconducting Nb resonator on STO at 1.1~GHz we find $Q$ around 800 at mK temperatures due to strong suppression of losses below 1~K. 

\begin{figure}
	\centering
	\includegraphics[width=\columnwidth]{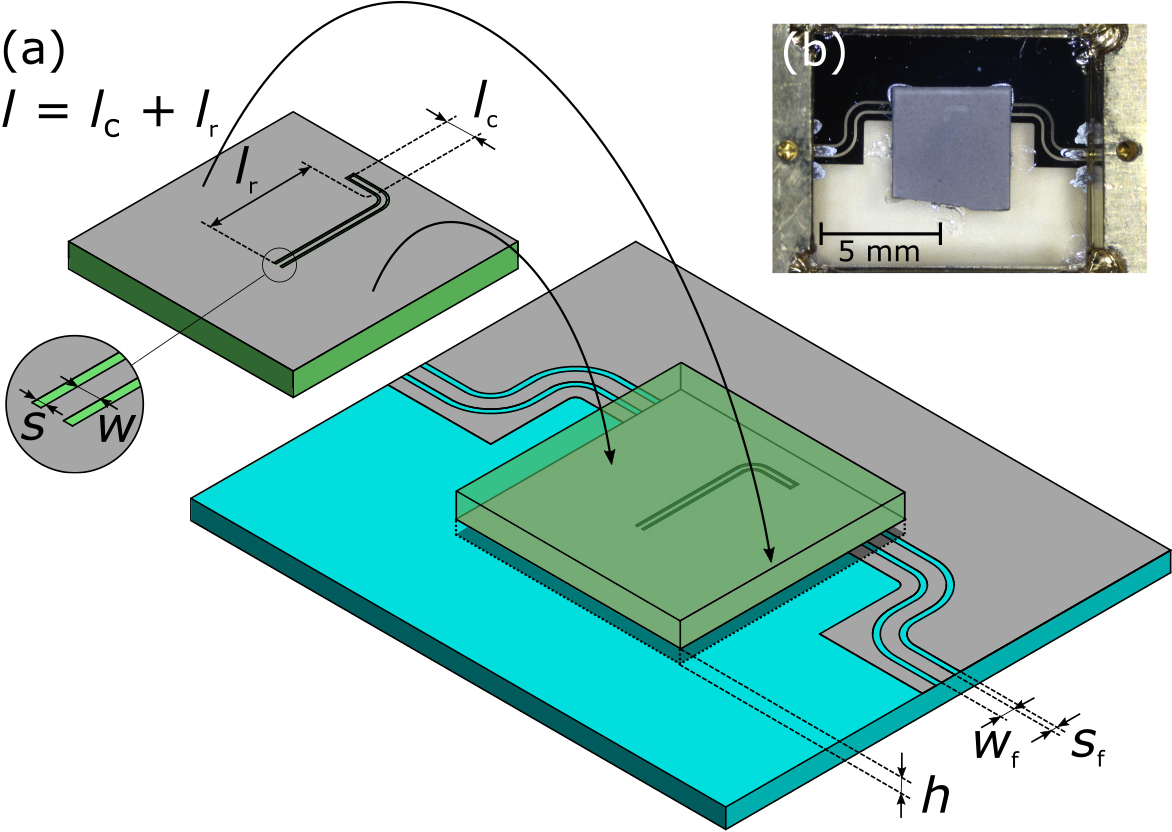}
	\caption{(a) Schematic sketch of the distant-flip-chip arrangement. The Nb film is shown in gray, the STO resonator substrate in green, and the Al$_2$O$_3$ feedline substrate in light blue. The resonator is flipped and centrally placed over the feedline with the coupling length $l_\text{c}$ at a distance $h$. (b) Photograph of the mounted chip where the parallel arm of the resonator above the feedline can be seen through the slightly transparent STO sample.}
	\label{Fig:Distant_flip-chip}
\end{figure}

We fabricated superconducting Nb coplanar $\lambda/4$-resonators on STO, as schematically shown in Fig.~\ref{Fig:Distant_flip-chip}(a). 
The main design idea is that the coplanar resonator is coupled to the microwave instrumentation via a coplanar 50~$\Omega$ feedline on a separate sapphire substrate, and that the resonator and feedline chips are mounted in a \lq distant-flip-chip\rq{} geometry \cite{Wendel2019}, i.e.\ the surfaces of the chips with the coplanar structures face each other, but they are held at some distance $h$. This way, the broadband (off-resonant) transmission of the microwave feedline is only weakly affected by the large dielectric constant of the STO, as it is held at a distance from the feedline and partially shielded by the \SI{300}{\nano\metre} thick superconducting Nb film. The strength of the on-resonance absorption of the resonator into the feedline signal can be tuned by design parameters: we chose inductive coupling between the feedline and a parallel section of length $l_\text{c}$ at the open end of the $\lambda/4$-resonator. The presented data are obtained with a resonator of total length $l=\SI{3200}{\micro\meter}$, coupling length $l_\text{c}=\SI{600}{\micro\meter}$, and distance $h \approx \SI{60}{\micro\meter}$.
The lateral dimensions are center conductor width $W = \SI{100}{\micro\meter}$ and center-strip-to-ground distance $S = \SI{40}{\micro\meter}$ for the resonator, while the feedline, also fabricated in Nb, has $W_\text{f} = \SI{300}{\micro\meter}$ and $S_\text{f} = \SI{120}{\micro\meter}$.
As shown in Fig.~\ref{Fig:Distant_flip-chip}(b), the \SI[product-units=power]{5x5}{\mm} STO (100) chip is slightly transparent, and thus allows optical alignment of the feedline and the coupling leg of the resonator. 
The STO chip is adjusted at distance $h$ using small amounts of vacuum grease at the four corners and then fixed with Fixogum.
The two-chip assembly is mounted in a brass sample box that incorporates coaxial feedthroughs, which are silver-pasted to the feedline. Microwave experiments are performed with a vector network analyzer (VNA) and for the mK experiments with an additional room-temperature amplifier, and the device is cooled down  in a $^4$He cryostat with variable temperature insert (VTI) for temperatures down to 1.4~K or in a $^3$He/$^4$He dilution refrigerator with base temperature 20~mK \cite{Thiemann2018a,Wiemann2015}. 

\begin{figure}
	\centering
	\includegraphics[width=\columnwidth]{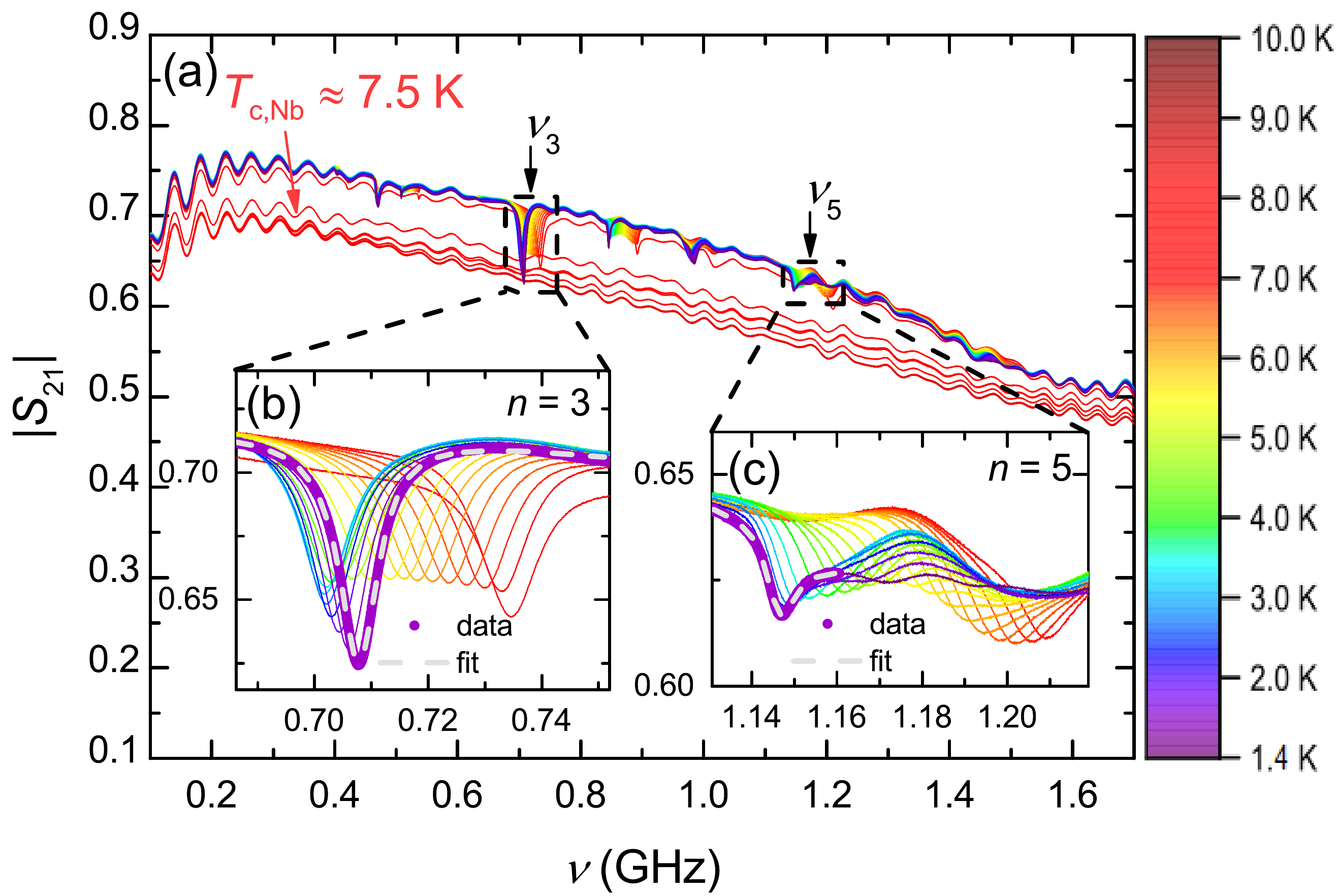}
	\caption{(a) Transmission spectra of the feedline coupled to a $\lambda/4$-resonator on an STO substrate, for temperatures from $T = \SI{1.4}{\K}$ to $T = \SI{10}{\K}$. Below the critical temperature $T_\textrm{c,Nb} = 7.5 \mathrm{K}$ the Nb resonator is superconducting, and temperature-dependent resonances are observed. (b,c) Resonant absorption, with a complex Lorentzian fit (dashed) for the $\SI{1.4}{\K}$ data, of (b) the resonator mode $n=3$ and (c) $n=5$.}
	\label{Fig:Alle_Moden}
\end{figure}

Fig.~\ref{Fig:Alle_Moden}(a) shows measured transmission spectra in the temperature range from about $T = \SI{1.4}{\K}$ to $T = \SI{10}{\K}$ and up to a frequency of around \SI{1.6}{\GHz}. 
At the highest temperatures, the spectrum exhibits an oscillatory pattern (due to undesired standing waves in the overall transmission line) and a continuous decrease towards higher frequencies (due to skin-depth-related losses). Upon cooling below the critical temperature $T_\textrm{c,Nb} \approx \SI{7.5}{\K}$ of the Nb,\cite{Farag2019} the feedline becomes superconducting, and the resulting reduced losses lead to increased broadband transmission \cite{Clauss2013}.
At the same time, well-defined absorptive resonances develop and become more pronounced upon further cooling. 
For a coplanar $\lambda/4$-resonator, the harmonic resonance frequencies follow
\begin{equation}
	\nu_{n} = \frac{nc}{4 l \sqrt{\epsilon_\text{eff}}}
	\label{eq:nu_vs_epsilon}
\end{equation}
with $n = 1,3,5,...$ the number of the mode, $c$ the vacuum speed of light, $l$ the complete resonator length and $\epsilon_\text{eff}$ the effective dielectric constant of the coplanar line.
Amongst the resonances in Fig.~\ref{Fig:Alle_Moden}(a), we identify the two pronounced ones near 0.71~GHz and 1.15~GHz as the expected $n = 3$ and $n = 5$ modes. Their temperature evolution is shown in detail in Figs.~\ref{Fig:Alle_Moden}(b) and (c), respectively.
Clearly, these resonances become sharper and shift in frequency upon cooling, and they can be properly described with Lorentzian fits (dashed lines) to obtain resonance frequency $\nu_{n}$ and quality factor $Q_{n} = \frac{\nu_{n}}{\Delta \nu_{n}}$ with $\Delta \nu_{n}$ the resonance linewidth \cite{CommentBackground}.
These data show that superconducting $\lambda/4$-resonators directly on STO for frequencies exceeding \SI{1}{GHz} can be realized also for cryogenic temperatures deep in the quantum-paraelectric regime of STO. 
At the same time, the spectra in Fig.~\ref{Fig:Alle_Moden}(a) evidence the experimental challenges related to the bulk STO: several resonances in addition to the desired $n = 3$ and $n = 5$ modes show up, which we assign as three-dimensional modes within the brass box or parasitic modes within the waveguide structure, all of which are affected by the presence of bulk STO. Also, the expected fundamental resonance near $\nu_{1} \approx \SI{0.24}{GHz}$ appears to be absent in the data.

\begin{figure}
	\centering
	\includegraphics[width=\columnwidth]{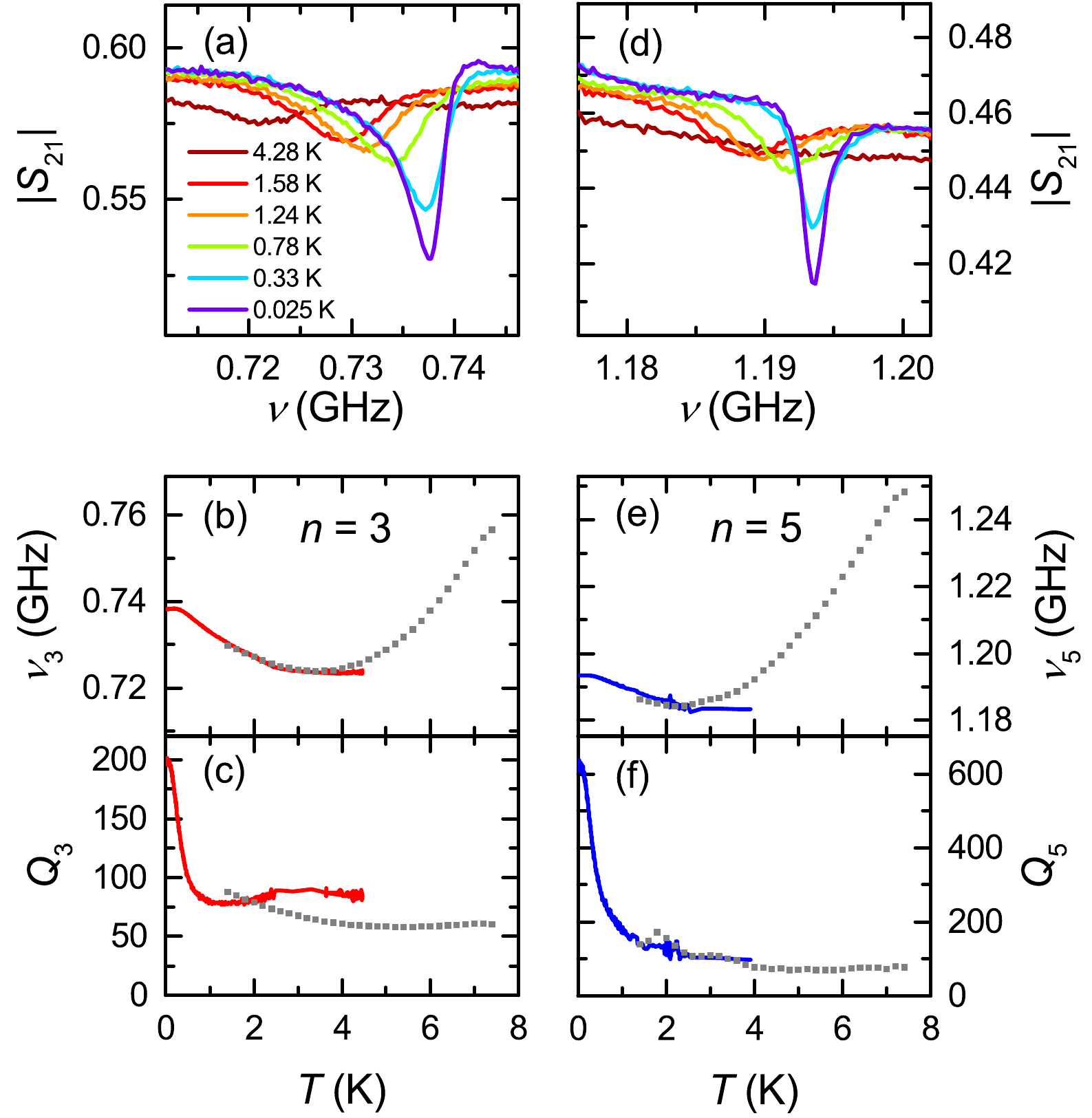}
	\caption{(a,d) Temperature-dependent spectra from $T = \SI{25}{mK}$ to $T = \SI{4.28}{K}$ for the modes (a) $n=3$ and (d) $n=5$. (b,e) Resonance frequency $\nu_n$ in dependence of the temperature from $T = \SI{25}{mK}$ to $T = \SI{7.5}{K}$ for (b) $n=3$ and (e) $n=5$. Frequencies of the high temperature measurement have been shifted for comparability \cite{footnoteshift_1}. (c,f) Quality factor $Q$ in dependence of the temperature from $T = \SI{25}{mK}$ to $T = \SI{7.5}{K}$ for (c) $n=3$ and (f) $n=5$. VTI data have been acquired at a microwave power of \SI{-30}{dBm}, \SI{}{mK} data at \SI{-40}{dBm}.}
	\label{fig:mischer}
\end{figure}

Figs.~\ref{fig:mischer}(a) and (d) show how the two modes $n = 3$ and $n = 5$ evolve further, at mK temperatures in the dilution refrigerator. 
These data immediately show a key result, namely that the resonances sharpen substantially when cooled well below 1~K, indicating a strong suppression of losses. 
For STO being a quantum paraelectric with Curie temperature around 30~K \cite{Neville1972,Mueller1979,Viana1994}, one would expect only very weak temperature dependence of $\hat{\epsilon}$ below 1~K, in accordance with previous sub-Kelvin studies of $\epsilon_1$ \cite{Mueller1979,Viana1994,Rowley2014}, while $\epsilon_2$ has not been reported for such low temperatures. Our observation of strong temperature dependence in $Q$ (and correspondingly in $\epsilon_2$ of STO) thus comes rather unexpectedly.

Figs.~\ref{fig:mischer}(b,e) and (d,f) show frequency $\nu_{n}$ and quality factor $Q_n$, respectively, for the resonator mode $n=3$ and $n=5$ in dependence of temperature, from $T \approx \SI{22}{mK}$ to $T \approx \SI{8}{K}$. 
There are two factors that cause shifts of $\nu_{n}$, namely the temperature dependence in $\epsilon_1$ of STO and the temperature dependence in the London penetration depth $\lambda_\text{L}$ of the Nb conductors. The latter affects the characteristic impedance of the coplanar line, and leads to an increase of $\nu_{n}$ with decreasing temperature and has very weak temperature dependence below $T_\text{c,Nb}/2$, i.e.\ below 4~K. In our present data, these Nb-related effects are completely overwhelmed by the temperature-dependent $\epsilon_1$ of STO, and thus we can evaluate $\epsilon_1(T)$ from the measured $\nu_{n}(T)$ via Eq.\ (\ref{eq:nu_vs_epsilon}) and a model that relates $\epsilon$ of the STO to the $\epsilon_\text{eff}$ of the resonator.

\begin{figure}
	\centering
	\includegraphics[width=\columnwidth]{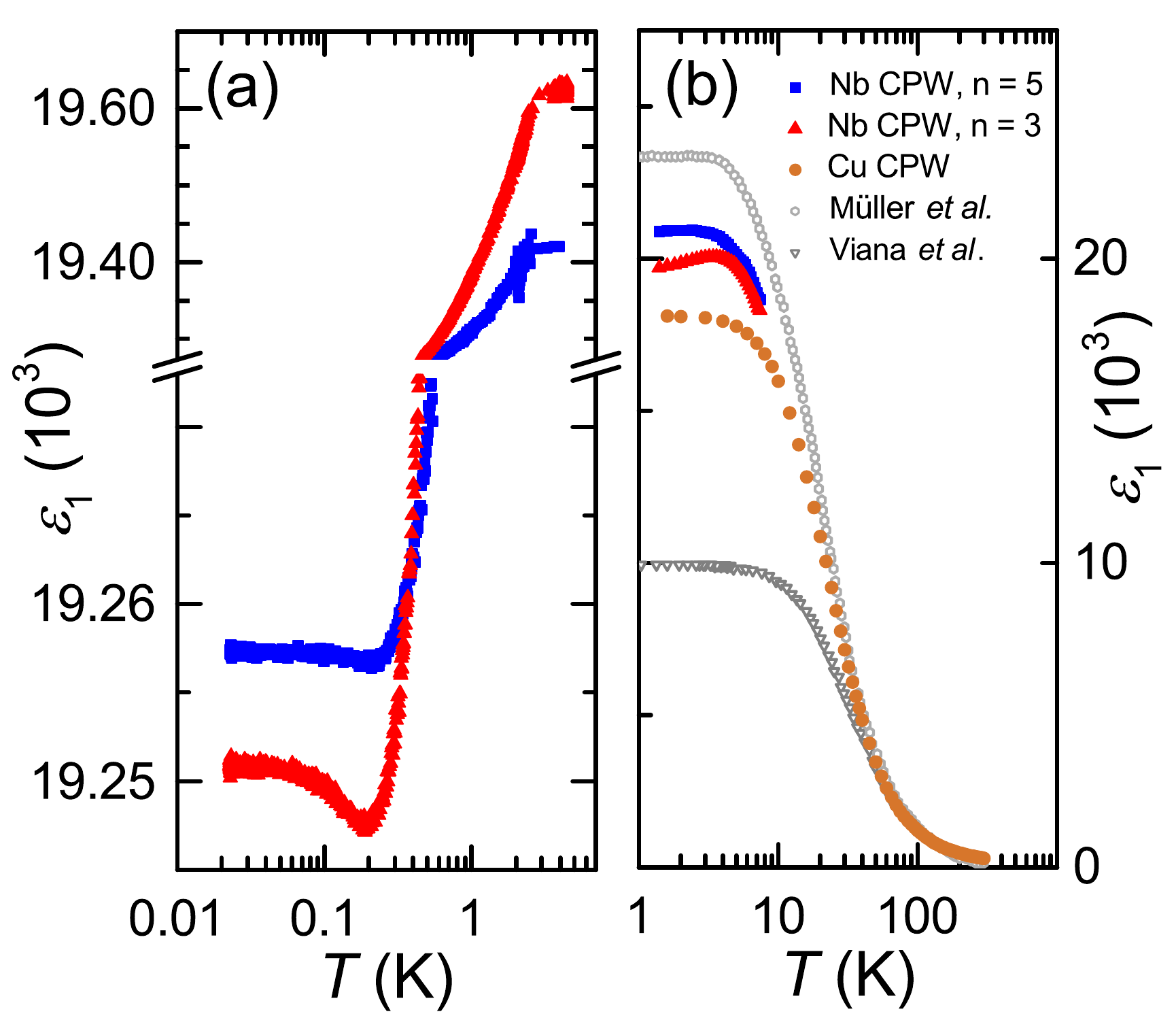}
	\caption{(a) Dielectric constant $\epsilon_1$ of STO obtained for resonator modes $n=3$ ($\nu = \SI{0.73}{GHz}$) and $n=5$ ($\nu = \SI{1.19}{GHz}$) for $T < \SI{1}{K}$. Data for $n=3$ has been shifted for comparability \cite{footnoteshift_2}. (b) Dielectric constant $\epsilon_1$ of STO obtained with different resonators from this work and comparison to literature \cite{Viana1994,Mueller1979} for temperatures from $T = \SI{1}{K}$ to $T = \SI{300}{K}$. VTI data have been acquired at a microwave power of \SI{-30}{dBm}, \SI{}{mK} data at \SI{-40}{dBm}.}
	\label{fig:eps1_alle}
\end{figure}

The resulting $\epsilon_1(T)$ are shown in Fig.\ \ref{fig:eps1_alle}(a) for the dilution-refrigerator measurements and in Fig.\ \ref{fig:eps1_alle}(b) for the VTI measurements. The latter plot includes an additional data set that we obtained using a conventional metallic coplanar resonator \cite{Javaheri2016} on STO (with resonator frequency changing from \SI{0.25}{GHz} at 2~K to \SI{1.99}{GHz} at 300~K) and two data sets from literature \cite{Viana1994,Mueller1979} to show the overall evolution of $\epsilon_1(T)$ with temperature.

Our data display the wide maximum in $\epsilon_1(T)$ near 2~K that was previously addressed with much lower probing frequencies \cite{Fischer1985,Mueller1979,Viana1994,Rowley2014}. 
Our measurements extend to much lower temperatures, and there we additionally find a weak, but well-resolved minimum in $\epsilon_1(T)$ around 200~mK. Such a minimum in $\epsilon_1(T)$ has so far not been discussed in the literature, but data of a previous kHz study might hint at such a feature \cite{Viana1994}. While the $\epsilon_1(T)$ maximum near 2~K has also been investigated from a theoretical point of view \cite{Rowley2014,Palova2009,Conduit2010,Chandra2017}, in particular in the context of quantum criticality, our data now suggest that an additional, even lower energy scale is present in STO.

Yet more dramatic is the sub-K behavior of $\epsilon_2(T)$. If the losses of a microwave resonator are caused solely by dielectric behavior, then the resonator $Q$ directly relates to $\tan \delta$ of the dielectric. In our particular device geometry we cannot rule out that the absolute value of $Q$ also contains loss contributions other than dielectric, in particular due to the coupling to the feedline circuitry. Therefore we discuss the losses in terms of the experimental quantity $Q$, but since losses due to coupling or in the superconducting Nb only very weakly depend on temperature below 1~K, it is clear that the dramatic $Q$ increase towards lowest temperatures in Figs. \ref{fig:mischer}(c) and (f) is caused by $\epsilon_2(T)$ of the STO.

\begin{figure}
	\centering
	\includegraphics[width=\columnwidth]{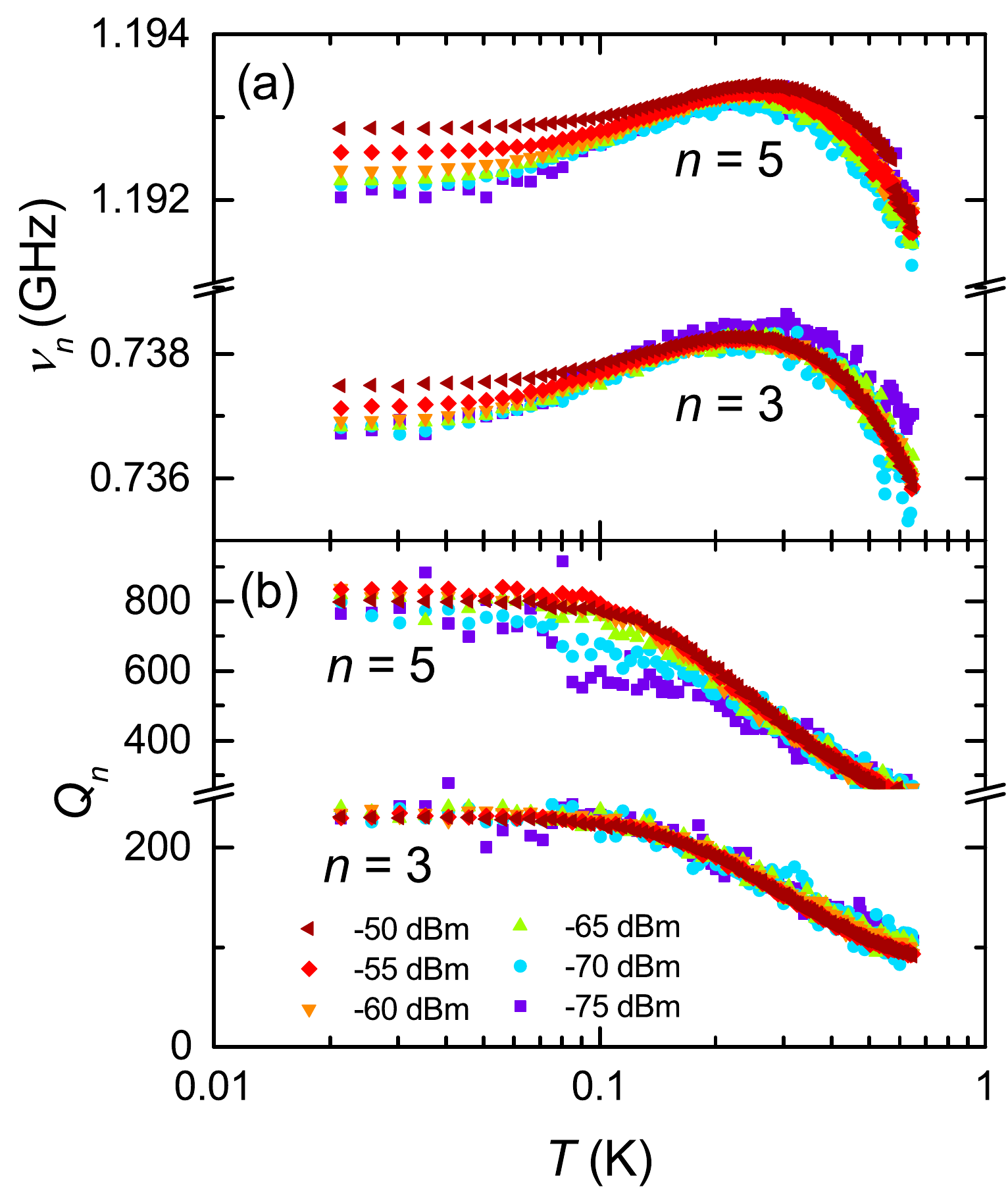}
	\caption{(a) Resonance frequency $\nu_n$ and (b) quality factor $Q_n$ for $n=3$ and $n=5$ in dependence of the temperature $T$ (measured at the mixing chamber), for different input powers $P$ ranging from $\SI{-75}{dBm}$ to $\SI{-50}{dBm}$.}
	\label{fig:mischer_power}
\end{figure}

To investigate this phenomenon further, we have performed measurements as a function of temperature and applied microwave power, a common approach to identify possible microwave heating as well as to distinguish certain loss mechanisms, in particular the role of two-level systems \cite{Barends2007,OConnell2008,Barends2010}.
The obtained data for $\nu_{n}$ and $Q_{n}$, for the same two modes as before, are displayed in Fig.\ \ref{fig:mischer_power} and show the same qualitative features already discussed: upon cooling below 1~K, $\nu_{n}$ reaches a maximum in the temperature range 0.2~K to 0.3~K and then decreases towards the lowest temperatures achieved. $Q$ continuously increases upon cooling, arguably reaching a plateau below 100~mK. The highest observed $Q$ for the $n=5$ mode approaches 840.

The characteristic temperature dependence of $\nu_{n}$ allows us to address possible undesired microwave heating: above 200~mK, the $\nu_{n}$ do not vary within experimental uncertainty when the power supplied by the VNA is increased from -75~dBm to -50~dBm, thus indicating absence of microwave heating in this regime. Below 100~mK, on the other hand, we see indications for heating, as spreading of the $\nu_{n}(T)$ data sets for different applied powers.
Here one can estimate that for our highest input power of -50~dBm and at lowest nominal temperature of 21~mK the actual sample temperature is around 100~mK, but for the lowest applied power of \SI{-75}{dBm} any residual heating will not increase the actual sample temperature above 40~mK. This consistent overall picture concerning microwave heating of the sample reinforces the identification of the peculiar $\nu_{n}(T)$ evolution at mK temperatures as due to intrinsic properties of $\epsilon_1(T)$ of STO.

The observed $Q_n$ for each applied power appears temperature independent below 100~mK, and thus these data sets will not be affected by the heating identified in the $\nu_{n}$ data. $Q_5$ seems to be reduced with decreasing power, which is consistent with fluctuating two-level systems as dominant loss channel. $Q_3$, in contrast, does not seem to be affected by the applied power in the range of this experiment. This might indicate that even the lowest applied power already saturates the two-level fluctuations, but this would be surprising considering that this is not the case for the $n=5$ mode at the same power despite its substantially higher $Q$. Here future experiments should also address yet lower input powers, which will be particularly relevant for any possible GHz quantum devices on STO.
Davidovikj \textit{et al.}\ also investigated the radio frequency (rf) response of STO as a function of applied power, and they found a strong increase of losses with increasing rf power \cite{Davidovikj2017}. This contrast to our observations is not surprising, since we operate at substantially lower temperatures, at higher frequencies, and at much lower applied powers, and thus we are much more susceptible to two-level systems.

In conclusion, we realized superconducting coplanar Nb resonators on STO substrates in a distant-flip-chip geometry operating at frequencies exceeding 1~GHz at cryogenic temperatures, which substantially extend the accessible frequency range for microwave circuits on quantum-paraelectric bulk STO.
Our experiments reveal new and unexpected dielectric behavior of STO at mK temperatures and GHz frequencies: at temperatures well below the previously investigated weak $\epsilon_1(T)$ maximum around 2~K we find non-monotonous behavior with an even weaker minimum in the temperature range 0.2-0.3~K. Even more remarkable is the temperature dependence of the dielectric losses, which reduce dramatically at temperatures below 1~K, leading to a resonator quality factor beyond 800 for a resonance near 1.2~GHz. These results challenge our understanding of the cryogenic dielectric properties of STO, and they call for additional efforts from both theory and experiment. Clearly our experiments can be improved further from a technical point of view, in particular concerning yet higher frequencies and the coupling to the microwave feedline. Even more intriguing are materials perspectives regarding the STO. With the underlying mechanism for the reduction of dielectric losses unclear, yet higher $Q$ seem achievable. The observed dramatic changes bring to mind the extreme sensitivity of the low-temperature dielectric properties of STO to parameters such as chemical and isotope composition or strain \cite{Rowley2014,Bednorz1984,Stucky2016}. Exploration of this abundant material phase space should also include more detailed consideration of the probing frequency for dielectric measurements. Interpretation of our results would profit from further lower-frequency studies in the same temperature range \cite{Mueller1979,Viana1994,Fischer1985,Rowley2014,Grams2014}, and extensions to higher frequencies are also desired \cite{Thiemann2018b,Hafner2014}, in particular considering the role of the well-studied low-lying soft mode for the dielectric properties of STO \cite{Kamaras1995,Yamanaka2000}.
Our experiment thus holds new perspectives for oxide quantum devices, e.g.\ based on superconducting LAO/STO at GHz frequencies, and at the same time reveals new facets of the enigmatic low-temperature properties of STO.

We thank G. Untereiner, A. Farag, and M. Ubl for their support with resonator preparation and M. Dressel, H. Boschker, and J. Hemberger for helpful discussions. We acknowledge financial support by the DFG.

V. T. E. and N. G. E. contributed equally to this work.


\begin{thebibliography}{1}
\bibitem{CommentBackground} Before Lorentzian fitting, the standing wave pattern of the broadband background is removed by subtracting the spectrum at $T = \SI{10}{\K}$, well above $T_\text{c,Nb}$, without the resonances.
\bibitem{footnoteshift_1} As the measurements for the two different temperature ranges were performed in different setups, the mounting of the resonator chip above the feedline differs slightly, resulting in a small shift of the resonance frequency values. The high-temperature resonance frequency values have therefore been shifted up by 0.022\,GHz ($n=3$) and 0.04\,GHz ($n=5$) to match the mK-temperature data.
\bibitem{footnoteshift_2} The $\epsilon_1$ data obtained with $n=3$ has been shifted up by a value of 765 for simpler comparison with the data obtained with $n=5$.
\end{thebibliography}
\end{document}